\documentclass[useAMS,usenatbib,fleqn]{mn2e}       
\usepackage{graphicx}
\usepackage{amssymb}
\usepackage{amsmath}
\usepackage{aas_macros}
\usepackage{deluxetable}
\usepackage{lscape}
\usepackage{subfigure}
\usepackage{hyperref}

\begin{document}

\title[The Eridanus Filaments]{The Origin of Ionized Filaments Within the Orion--Eridanus Superbubble}

\author[A. Pon et al.] {Andy Pon, $^{1,2,3}$ Doug Johnstone, $^{4,3,2}$ John Bally, $^{5}$ \& Carl Heiles $^6$ \\
$^1$School of Physics and Astronomy, University of Leeds, Leeds LS2 9JT, UK\\
$^2$Department of Physics and Astronomy, University of Victoria, PO Box 3055 STN CSC, Victoria, BC V8W 3P6, Canada\\
$^3$NRC-Herzberg Institute of Astrophysics, 5071 West Saanich Road, Victoria, BC V9E 2E7, Canada\\
$^4$Joint Astronomy Centre, 660 North A'ohoku Place, University Park, Hilo, HI 96720, USA\\
$^5$Department of Astrophysical and Planetary Sciences, University of Colorado, UCB 389 CASA, Boulder, CO 80389-0389, USA\\
$^6$Astronomy Department, University of California, 601 Campbell Hall 3411, Berkeley, CA 94720-3411, USA}

\maketitle

\begin{abstract}
The Orion--Eridanus superbubble, formed by the nearby Orion high mass star-forming region, contains multiple bright H$\alpha$ filaments on the Eridanus side of the superbubble. We examine the implications of the H$\alpha$ brightnesses and sizes of these filaments, the Eridanus filaments. We find that either the filaments must be highly elongated along the line of sight or they cannot be equilibrium structures illuminated solely by the Orion star-forming region. The Eridanus filaments may, instead, have formed when the Orion--Eridanus superbubble encountered and compressed a pre-existing, ionized gas cloud, such that the filaments are now out of equilibrium and slowly recombining.
\end{abstract}


\section{INTRODUCTION}
\label{intro}
			 	 	 
	 The closest high-mass star-forming region to the Sun that is currently forming massive stars is the Orion star-forming region, which is located at a distance of 400 pc from the Sun \citep{Hirota07, Menten07, Sandstrom07}. The Orion star-forming region is surrounded by a highly elongated superbubble, with dimensions of 20$^\circ \times 45^\circ$ as seen in H$\alpha$ emission \citep{Bally08}, that is referred to as the Orion--Eridanus superbubble. 
	 
\begin{figure}
   \centering
   \includegraphics[width=3in]{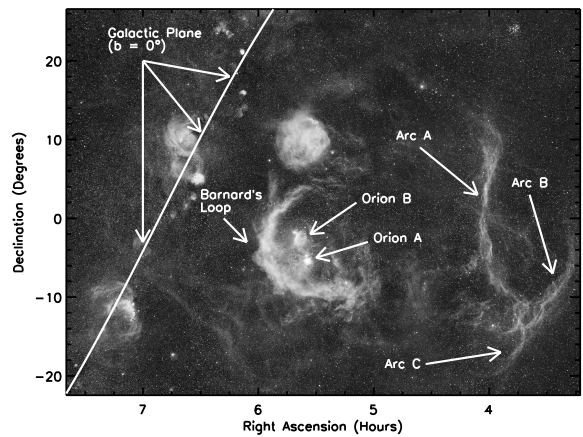}
   \caption{Orion--Eridanus superbubble as seen in H$\alpha$. Labels for the various major components of the bubble have been added to the image from \citet{DiCicco09}. Arcs A, B, and C are collectively referred to as the Eridanus filaments.}
   \label{fig:dicicco}
\end{figure}

	The Eridanus side of the superbubble contains a very prominent hook-shaped H$\alpha$ feature that was first discovered on H$\alpha$ images and Palomar Observatory Sky Survey plates by \citet{Meaburn65, Meaburn67}. \citet{Johnson78} break this hook into three separate arcs, all of which are labelled in Fig. \ref{fig:dicicco}. Arc A is the eastern half of the hook, Arc B is the western half of the hook, and Arc C is the southern extension of the hook. For the remainder of this paper, we will refer to these three arcs collectively as the Eridanus filaments. In this paper, all references to north or south refer to increasing or decreasing declination and references to east or west refer to increasing or decreasing right ascension, unless otherwise specified. 
	
	The Eridanus filaments are clearly brighter, in H$\alpha$, than the superbubble wall in their vicinity. They are also remarkably bright, with H$\alpha$ intensities of the order of 25 Rayleighs, given that they are located almost 200 pc from the Orion star-forming region \citep{Haffner03}. In this paper, we examine the implications of the H$\alpha$ brightnesses and sizes of these filaments, with a particular eye towards whether the filaments are consistent with being in ionization equilibrium with the Orion star-forming region.
	
	The H$\alpha$ emission from a radiatively excited region is proportional to the amount of ionizing flux absorbed by that region, because the H$\alpha$ emission is generated from the recombination of bare protons and electrons. As such, information on the density and geometry of the emitting region, as well as the strength of the ionizing radiation field, can be obtained from H$\alpha$ intensities and distributions (e.g. \citealt{Reynolds79, Heiles99}). Strong dynamical events can also imprint their signatures into the H$\alpha$ emission from a gas cloud, such that ionization modelling can provide a window into the dynamical history of the H$\alpha$ emitting region. 
	
	In Section \ref{properties}, we review and derive the general properties of the Eridanus filaments. In Section \ref{equilibrium}, we examine the criteria required for the Eridanus filaments to be in ionization equilibrium with the Orion star-forming region, while in Section \ref{alternative models}, we examine alternative possibilities for the ionization state of the filaments. We briefly discuss the possible origins of the filaments in Section \ref{origin} and we summarize our primary findings in Section \ref{conclusions}. 
		
\section{GENERAL PROPERTIES}
\label{properties}

	The distances to the Eridanus filaments are not well known. Studies of interstellar absorption features in stellar spectra towards the Eridanus half of the Orion--Eridanus superbubble reveal a wall of gas moving towards the Sun at a distance of approximately 180 pc, which has often been interpreted as the near wall of the superbubble \citep{Guo95, Burrows96,Welsh05}. If the filaments are associated with the near side of the Orion--Eridanus superbubble, they would thus be only 180 pc distant. There is also some evidence, see Appendix \ref{Arc A}, that Arc A and the back side of the Orion--Eridanus superbubble are located at a distance greater than 500 pc \citep{Boumis01, Welsh05, Ryu06}.

	The Eridanus features have been detected in numerous H$\alpha$ surveys, including that by the Wisconsin H-Alpha Mapper (WHAM; \citealt{Haffner03}). All values for the intensities of the filaments quoted below will be from the WHAM survey, if not otherwise stated. 
	
	Arc A is the brightest of the Eridanus filaments, with a peak intensity of 70 Rayleighs and a typical intensity closer to 25 Rayleighs. Arcs B and C have slightly lower typical intensities of 15 and 10 Rayleighs, respectively. To the west of Arc B, the H$\alpha$ intensity drops to 1 Rayleigh, whereas the intensity is closer to 5 Rayleigh throughout the interior of the Orion--Eridanus superbubble to the east of Arc A. 
	
	It is slightly odd that while the bubble wall shows clear signs of limb brightening, with the north and south edges of the bubble wall appearing more prominently in Fig. \ref{fig:dicicco}, the filaments do not show any significant limb brightening. That is, the north and south ends of the filaments do not appear to be significantly brighter than the middle of the filaments. 
	
	There are numerous H {\sc i} features that correlate well with integrated H$\alpha$ emission features (e.g, \citealt{Hartmann97, Arnal00, Bajaja05, Kalberla05}). In particular, H {\sc i} filaments are detected just to the west of Arcs A, B, and C. The H {\sc i} filament closest to Arc B lies approximately 3$^\circ$ to the west of the arc (e.g., \citealt{Verschuur92}), which corresponds to a physical separation between 10 and 25 pc for distances of the filaments from the Sun between 200 and 500 pc. Zeeman splitting measurements indicate that a partially ionized region, with a line-of-sight magnetic field of the order of 10 $\mu$Gauss, lies between Arcs B and C and the nearest H {\sc i} filament \citep{Heiles89}. This spatial morphology is consistent with the H$\alpha$ coming from the ionized interior edge of a shell and the H {\sc i} coming from the neutral exterior of the shell that is shielded from the ionizing photons of the Orion star-forming region by the inner regions of the shell.

\subsection{Densities and Depths}
\label{densities and depths}	

	The observable emission measure of the H$\alpha$ transition, EM, is dependent upon both the number density of the emitting gas, $n$, and the line-of-sight depth of the emitting gas, $R$, via
\begin{equation}
\mbox{EM} = \int n^2 \mbox{d}R.
\label{eqn:EM}
\end{equation}
Based upon the observed intensity of the filaments, it is thus possible to either calculate the depth of the filaments, given an estimate of the density of the filaments, or to calculate the density of the filaments, given an estimate of their depths. 
	
	Since all of the Eridanus filaments are associated with both H {\sc i} and H$\alpha$ emission, it appears that the filaments all contain ionization fronts. Therefore, the filaments should have a temperature of approximately 8000 K \citep{Basu99}, for which an H$\alpha$ intensity of 1 Rayleigh corresponds to an emission measure of 2.25 pc cm$^{-6}$ (e.g., \citealt{Haffner03}). 	
		
	We assume that the gas pressure in the filaments is equal to the gas pressure within the Orion--Eridanus superbubble, although \citet{BisnovatyiKogan95} suggest that the pressure within a superbubble's wall may be twice that of the interior of the bubble. We adopt a pressure range of 1-$5 \times 10^{4}$ K cm$^{-3}$ for the interior of the Orion--Eridanus superbubble, and thus the Eridanus filaments, based upon previous observations \citep{Burrows93, Guo95, Burrows96}. For this pressure range and a temperature of 8000 K, the number density of the filaments is of the order of 1-6 cm$^{-3}$. 
	
	Alternatively, if the expansion velocity of the superbubble is taken to be 15 km s$^{-1}$ \citep{Reynolds79} and it is assumed that the expansion provides a ram pressure on the interior of the bubble to match the thermal pressure in the bubble, the filaments would have a density of the order of 1 cm$^{-3}$. 
	
	These densities are consistent with the densities derived by \citet{Reynolds79} and \citet{Heiles89}. \citet{Reynolds79} derive an electron density of $1.1 \,(D / 400\mbox{ pc})^{0.5}$ cm$^{-3}$ for the Eridanus filaments while \citet{Heiles89} derive a density of 6 cm$^{-3}$ for the H {\sc i} filament tracing Arcs B and C. If the temperature of the filaments is slightly lower than 8000 K, as suggested by \citet{Heiles00}, or if the filament pressure is twice the interior pressure, as predicted by \citet{BisnovatyiKogan95}, the above-calculated densities would increase by approximately a factor of 2. Conversely, if there is a pressure gradient within the bubble, such that the bubble pressure is lower near the filaments \citep{Burrows96}, the filament densities could be lower by a factor of a few. 
	
	Since the typical observed intensity of the filaments is 15-25 Rayleighs \citep{Haffner03}, the depth of the ionized material in these filaments must be of the order of 1-50 pc, given the density range of 1-6 cm$^{-3}$. 
	
	Alternatively, if we make the assumption that the depths of the filaments are approximately equal to their widths, we can derive a density range for the filaments. Both Arcs A and B have angular widths of approximately 1$^\circ$.5, which, if the arcs are 180 pc distant, would correspond to physical widths of 5 pc. If the arcs are 500 pc distant, their widths would be closer to 13 pc. For a depth of 5 pc, the number density required to produce the observed emission measure is of the order of 3 cm$^{-3}$, while for a depth of 13 pc, the required density is between 1.5 and 2 cm$^{-3}$. These densities are consistent with the densities estimated from the interior pressure of the bubble. 

	Under the above assumption of cylindrical geometry for the filaments, such that their depths are equal to their widths, the column densities of these filaments along our line of sight would be between 4 and 8 $\times 10^{19}$ cm$^{-2}$. This corresponds to visual extinctions (A$_V$) between 0.01 and 0.05. Since this is only the column density of ionized gas in the filaments, it is somewhat expected that these columns are slightly lower than the column density of a few times 10$^{20}$ derived towards the southern half of Arc A from other methods \citep{Heiles89,Burrows93, Guo95,Snowden95Burrows}. Since Arc A is roughly 25$^\circ$ long and Arc B is 15$^\circ$ long, the total mass of ionized gas in the filaments is approximately $3 \times 10^{2}$ M$_\odot$  if the filaments are 180 pc distant, and $3 \times 10^{3}$ M$_\odot$ if the filaments are 500 pc distant. 

	The peak intensity of Arc A, 70 Rayleighs \citep{Haffner03}, requires an unrealistically large depth of 70 pc for a density of 1.5 cm$^{-3}$, but only requires a depth of 17.5 pc for a density of 3 cm$^{-3}$. At this higher density, the peak intensity would correspond to a column density of $2 \times 10^{20}$ cm$^{-2}$, which is still significantly smaller than the column density of $1.9 \times 10^{21}$ cm$^{-2}$ derived towards the northern half of Arc A by \citet{Heiles00}. The higher column density estimated by \citet{Heiles00} could be due to neutral hydrogen also being present in Arc A, as would be expected if the ionizing photons from the Orion star-forming region are fully trapped within Arc A. The much larger peak intensity of Barnard's Loop, 250 Rayleighs, is likely due to a further increase in the density of the loop, rather than being due to a very large line-of-sight depth.  

\section{IONIZATION EQUILIBRIUM MODELS}
\label{equilibrium}

	While there is some debate in the literature regarding whether Arcs A and B are part of the Orion--Eridanus superbubble (e.g., \citealt{Boumis01}, see also Appendix \ref{Arc A}), for this section, we make the assumption that the filaments are in ionization equilibrium with the Orion star-forming region. We will later drop this assumption in Section \ref{alternative models}

	Because H$\alpha$ emission is powered by the absorption of ionizing photons, H$\alpha$ intensity should be proportional to the energy of ionizing radiation absorbed. This proportionality, however, can be broken if there is another energy source powering the ionization of the emitting gas or if the emitting regions are not in ionization equilibrium. For a superbubble, the H$\alpha$ surface luminosity of the superbubble's wall should vary roughly with the flux of incident ionizing photons arriving from the ionizing source wherever the ionization front occurs within the bubble wall, since all of the incident ionizing photons are absorbed within the wall at these locations.. As such, the H$\alpha$ flux in these regions should vary inversely with the square of the distance from the ionizing source. For regions where the ionization front lies outside of the superbubble, the H$\alpha$ intensity should depend upon both the ionizing flux reaching the wall and the fraction of the ionizing flux that the wall captures. Thus, the H$\alpha$ brightness of the wall should show a discontinuity where the ionization front breaks out, with the H$\alpha$ intensity beyond the point where the ionization front breaks out being lower than that predicted from the inverse square distance dimming seen before the ionization front breaks out. 

	For the Eridanus filaments to be brighter in H$\alpha$ than the parts of the bubble wall closer to the Orion star-forming region, the superbubble walls near the filaments must not fully trap the ionizing photons incident upon them and the filaments must trap a greater fraction of the ionizing photons than the surrounding walls. The inability of the walls near the filaments to fully trap the ionizing photons incident upon them is consistent with the ionization front breaking out of the Orion--Eridanus superbubble's wall at the western ends of Barnard's Loop, where there is a significant decrease in H$\alpha$ intensity. It is, however, unclear where the extra mass in the filaments required to trap significantly more ionizing photons would have come from, as most models of superbubbles would predict relatively low gas masses at such large distances from the driving source \citep{Basu99}. The increase in brightness of the filaments relative to the adjacent bubble wall could, alternatively, just be due to an increase in the line-of-sight depth of the emitting gas in the filaments. In this case, the filaments would have a larger surface area over which to absorb energy from the Orion star-forming region, such that their emission could be greater than that of the surrounding bubble wall. 
	
	Proceeding with the assumption that the relative brightnesses of the Eridanus filaments and the surrounding bubble wall can be explained by variable fractions of ionizing photons absorbed or by changes in line-of-sight depth, the absolute intensity of the filaments must still be explained. If the filaments are in ionization equilibrium, there must not only be a large enough energy flux in ionizing photons incident upon the filaments to provide the power for the H$\alpha$ emission, but there must also be enough material within the filaments to absorb a large enough fraction of this incident ionizing photon energy. H {\sc i} emission associated with the Eridanus filaments is clearly detected, indicating that there is enough material in the filaments to fully absorb all of the incident ionizing photons from the Orion star-forming region. Thus, we concentrate in the following paragraphs on whether the Orion star-forming region has a large enough ionizing photon luminosity to explain the observed characteristics of the Eridanus filaments. 	

We introduce a geometric scaling parameter, $\gamma_r$, to relate the line-of-sight depth from the Sun through the ionized portions of the filaments, $R$, to the depth to which the ionizing photons from the Orion star-forming region have ionized the filaments as measured from the Orion star-forming region, $L$, via
\begin{equation}
R = \gamma_r L.
\label{eqn:gammar}
\end{equation}
For cylindrical symmetry, $\gamma_r = 1$ at the midpoint of the filaments. For the Eridanus filaments, we expect that $\gamma_r$ should be of order unity.

The depth to which ionizing photons can ionize material, $d$, is given by
\begin{equation}
d = \frac{\Phi_*}{4 \pi \, n^2 \, \alpha_b \, s^2},
\label{eqn:iondepth}
\end{equation}
where $\Phi_*$ is the total ionizing luminosity of the source, $n$ is the density of the gas, $\alpha_b$ is the recombination coefficient for hydrogen, and $s$ is the distance to the ionizing source. 

Under the assumption that the filaments have fully trapped the ionizing photons from the Orion star-forming region, such that $d = L$, combining and simplifying Equations \ref{eqn:EM} through \ref{eqn:iondepth} yields an expression for the ionizing luminosity of the Orion star-forming region in terms of the emission measure from the filaments as
\begin{equation}
\gamma_r \, \Phi_* = 4 \pi \, \mbox{EM} \, \alpha_b \, s^2 .
\label{eqn:gammarphi}
\end{equation}

Based upon the total H$\alpha$ emission from the Orion--Eridanus superbubble region, \citet{Reynolds79} calculate that the Orion star-forming region must have an ionizing luminosity of the order of $4 \times 10^{49}$ s$^{-1}$. \citet{Odell11} also independently determine the ionizing luminosity of the Orion star-forming region to be $1.9 \times 10^{49}$ s$^{-1}$, based upon the O-star models of \citet{Heap06} and the spectral classifications of stars in the Orion star-forming region by \citet{Goudis82}, and $2.5 \times 10^{49}$ s$^{-1}$ based upon the H$\beta$ brightness of Barnard's Loop. 

For a distance of 220 pc between the filaments and the Orion star-forming region, an ionizing luminosity of $4 \times 10^{49}$ s$^{-1}$, a geometric $\gamma_r$ value of 1, and a hydrogen recombination coefficient of $2.6 \times 10^{-13}$ cm$^3$ s$^{-1}$, the expected emission measure from the Eridanus filaments is only 9 pc cm$^{-6}$, which would correspond to 4 Rayleighs. This is a factor of 4-6 lower than the observed 15-25 Rayleigh intensity of the filaments. To obtain an H$\alpha$ intensity between 15 and 25 Rayleighs, the product of the ionizing luminosity of Orion and $\gamma_r$ must be increased by this factor of 4-6. It seems unlikely that the ionizing luminosity of Orion has been underestimated by this large of a factor such that if this ionization equilibrium model is correct, it is likely that $\gamma_r$ must be large.

By treating the filaments as portions of sheets, the value of $\gamma_r$ can be related to the geometry of the filaments. That is, the filaments have to be oriented such that cos($\phi$) / cos($\theta$) = $\gamma_r$, where $\theta$ is the angle between the surface normal of a filament and the line of sight between the filament and the Sun, and $\phi$ is the angle between the surface normal of the filament and the line of sight from the Orion star-forming region to the filament. For a $\gamma_r$ value of 4, $\theta$ must be at least 75$^\circ$, and is more likely to be close to 80$^\circ$. Kompaneets models of superbubbles \citep{Kompaneets60, Basu99} have been shown to well reproduce the shapes of superbubbles \citep{MacLow89,Basu99,Stil09}, and the best-fitting Kompaneets models to the Orion--Eridanus superbubble \citep{Pon14b} predict that $\theta$ should be between 5$^\circ$ and 30$^\circ$ if the filaments are part of the superbubble wall. 

While tangential sightlines to thin rings can produce significant limb brightening, this cannot explain the large $\gamma_r$ value required for the filaments since the filaments are roughly just as bright at their midpoints as at their northern and southern edges. At the midpoint of the filaments, there is no added depth due to a tangential sightline through the ring traced out by the filaments.

The H$\alpha$ brightness of the superbubble wall away from the Eridanus filaments is of the order of 5 Rayleighs, which is the intensity predicted for the currently accepted ionizing photon luminosity of the Orion star-forming region. Thus, a model in which the walls are in ionization equilibrium with the Orion star-forming region can explain the wall brightness. Alternatively, if the bubble wall only has a column density of $3 \times 10^{18}$ cm$^{-2}$, as suggested by \citet{Burrows93}, much of this H$\alpha$ emission may be coming from gas outside of the bubble that has been ionized by the photons passing through the bubble wall, rather than coming from the bubble wall itself. \citet{Odell11} also show that Barnard's Loop is consistent with being ionized by the bright stars of the Orion star-forming region. 

The depth to which ionizing photons penetrate can also be compared to the observed widths of the Eridanus filaments. As before, we introduce a geometric scaling parameter, $\gamma_t$, which relates the width of the filaments on the plane of the sky, $W$, to the depth to which the ionizing photons from the Orion star-forming region have ionized the filaments as measured from the Orion star-forming region via:
\begin{equation}
W = \gamma_t L.
\label{eqn:gammat}
\end{equation}
If the Eridanus filaments were a cylindrical ring, then $\gamma_t =1$. 

Under the assumption that the filaments fully trap the ionizing photons from the Orion star-forming region, Equation \ref{eqn:iondepth} can be re-written as
\begin{equation}
\gamma_t \, \Phi_* = 4 \pi \, n^2 \, \alpha_b \, s^2 \, W.
\label{eqn:gammatphi}
\end{equation}
Unlike in Equation \ref{eqn:gammarphi}, the above expression still maintains a dependence upon the density of the filaments. 

We choose to evaluate Equation \ref{eqn:gammatphi} with the density range derived from the bubble's interior pressure. We do not use the smaller density range for the filaments that we previously derived by assuming that the line-of-sight depth of the filaments is equal to their width because we do not want to presuppose the geometry of the filaments. If we were to use such a cylindrical geometry approximation, Equation \ref{eqn:gammatphi} would just reduce to Equation \ref{eqn:gammarphi}. For the density range of 1-6 cm$^{-3}$ and a range of filament widths from 5 to 13 pc, the value of $\gamma_t \, \Phi_*$ ranges from $2 \times 10^{49}$ s$^{-1}$, consistent with the estimated ionizing luminosity for the Orion star-forming region, all the way up to $2 \times 10^{51}$ s$^{-1}$, a factor of 100 larger than expected. 

While the observed widths of the filaments are consistent with the depth to which the ionizing photons from the Orion star-forming region are capable of penetrating through the filaments, it is only for the extreme edge of allowed parameter space that this consistency is achieved. That is, for agreement, the filament densities must be as low as possible and the filaments must be as close to the Sun as possible, such that the filament widths are as small as possible. In such a case, however, the filaments would still have a depth, $R$, roughly a factor of 5 larger than their widths in order to produce the required emission measure. In this case, the filaments would still have to be fairly highly elongated, edge-on sheets.

From a combination of dust continuum, H {\sc i}, and H$\alpha$ data, \citet{Heiles99} estimate that the electron density in Arc A is approximately 1 cm$^{-3}$ and find that Arc A must have a depth between 32 and 120 pc. Since the width of Arc A would be 14 pc at a distance of 400 pc, \citet{Heiles99} also conclude that Arc A is not a filament, but rather, an edge-on sheet. \citet{Heiles00}, however, apply this same technique to Barnard's Loop and find that Barnard's Loop must also have a depth of 160 pc. \citet{Heiles00} note that by modifying their assumed grain size distribution, they can reduce this depth by a factor of 4 and increase the electron density by a factor of 2 such that they conclude that the observed fluxes from Barnard's Loop should still be consistent with a filamentary model. Such a similar change to the assumed grain size distribution of \citet{Heiles99} may allow Arc A to also be consistent with a filamentary model.

	The presence of additional material within the superbubble, such as dust photoablated from the Orion molecular clouds, would absorb additional UV photons and cause the UV radiation field to decrease faster than predicted from geometric effects alone. Such material within the bubble would not emit significant H$\alpha$ flux, as the amount of H$\alpha$ intensity emitted per unit column density is much lower at 10$^6$ K, as in the interior of the bubble \citep{Burrows93}, than it is at 10$^4$ K, as in the bubble wall. Material within the bubble would reduce the UV flux reaching the Eridanus filaments, making it even more difficult to explain the Eridanus filaments as being in ionization equilibrium with the Orion star-forming region. Similarly, \citet{Heiles00} estimate the reddening towards one location in Arc A, based upon the relative observed intensities of H$\alpha$, [N {\sc ii}], and 2325 MHz radio continuum emission, and suggest that the unabsorbed H$\alpha$ intensity towards the arcs may be as much as a factor of 3.7 higher than observed. A larger H$\alpha$ intensity for the filaments would require even larger values of $\gamma_r \, \Phi_*$.

If the filaments were porous such that they were composed of small, denser pockets of gas, then large values of $\gamma_t$ could be obtained. Since the emission measure and ionization depth have the same dependence on density, decreasing the volume filling factor of the filaments, however, would have no impact on $\gamma_r$. 

\section{ALTERNATIVE MODELS}
\label{alternative models}

In Section \ref{equilibrium}, it was shown that if the Eridanus filaments are in ionization equilibrium and ionized by the Orion star-forming region, they would either have to be relatively edge-on sheets, inclined at an angle of about 80$^\circ$ to the plane of the sky, or the Orion star-forming region would have to have an ionizing luminosity much larger than previously estimated. It is somewhat unlikely that the ionizing luminosity of the Orion star-forming region has been underestimated by a factor of 5. While it is possible that the filaments are sheets that are nearly edge on over their entire lengths, such a configuration is highly constrained. As such, we also look for alternative models for the filaments where we drop the requirement that the filaments are in ionization equilibrium with the Orion star-forming region. 

	A secondary ionization source for the filaments, such as shocks, a recent UV flash, or the hot plasma filling the bubble \citep{Boumis01}, might account for the size and brightness of the filaments. While strong shocks are capable of ionizing hydrogen, \citet{Reynolds79} note, based on \citet{Cox72} and \citet{Raymond76}, that a 20-30 km s$^{-1}$ shock would not ionize enough hydrogen to produce the observed H$\alpha$ brightness of the Eridanus filaments. A higher [O {\sc iii}]-to-H$\alpha$ ratio than seen by \citet{Reynolds79} towards the superbubble would also be expected for ionization due to hard UV, soft X-rays, or cosmic rays \citep{Bergeron71}. Furthermore, there are no stars on the Eridanus side of the bubble that would have had enough energy to have formed a separate bubble which Arc A might be a part of \citep{Heiles76}. 

	Alternatively, the discrepancy between the observed thickness and depth of the Eridanus filaments and the depth to which ionizing photons from Orion can ionize material can be resolved if the Eridanus filaments are not equilibrium objects. Since the recombination rate of hydrogen is dependent upon the square of the density of the gas, a greater column of gas can be ionized by the same ionizing flux if the gas is at a lower density. If the Eridanus filaments were formed via the compression of a pre-existing, ionized gas cloud, the column of ionized gas within the final compressed filament would be larger than the column that could be currently ionized, and thus, the H$\alpha$ intensity from such a compressed filament would be larger than if the filament were in ionization equilibrium. A low-density gas cloud initially situated beyond the superbubble could have been ionized by the ionizing photons from the Orion star-forming region because the superbubble wall towards the Eridanus side of the Orion--Eridanus superbubble does not appear to fully trap ionizing photons. Since strong shocks can increase the density of a gas by up to a factor of 4, the shocks induced in a pre-existing cloud when the superbubble wall encounters the cloud could increase the emission coming from the cloud by a factor of 4, which is roughly what is needed to explain the observed brightness of the filaments. We thus suggest, as an alternative to the filaments being highly inclined sheets, that the Eridanus filaments might have formed from the compression of a pre-existing, ionized gas cloud by a strong shock and that the filaments are currently out of ionization equilibrium.
		
	For a hydrogen recombination coefficient of $2.6 \times 10^{-13}$ cm$^3$ s$^{-1}$, the recombination time is approximately 1.2 Myr $\left(n / \mbox{cm}^{-3}\right)^{-1}$ such that for a density of the order of a few cm$^{-3}$, the recombination time is just slightly less than 10$^6$ yr. While this time-scale is shorter than the few Myr age of the superbubble, estimated from the observed ages of Orion subgroups \citep{Brown94}, it is not unrealistically small for the Eridanus filaments to still be in a non-equilibrium phase. 
	
	\citet{Reynolds79} estimate that the line-of-sight expansion velocity of the Orion--Eridanus superbubble is 15 km s$^{-1}$, based upon observed line splitting. This line-of-sight velocity must be considerably less than the average total expansion velocity of the superbubble, as an average speed of 35 km s$^{-1}$ is required for the superbubble to have expanded to its full 300 pc length, assuming an upper limit of 8 Myr for the age of the bubble based upon the time since the formation of Orion OB1b \citep{Brown94}. It is quite possible that the superbubble could be a factor of 2 or more younger, thereby requiring an average expansion speed greater than 50 km s$^{-1}$. Given a temperature of the order of 10$^6$ K, the internal sound speed of the superbubble should also be of the order of 100 km s$^{-1}$. 

	If the filaments are shock-compressed features, then their original sizes must have been at least roughly a factor of 4 larger, given that strong shocks induce a density increase of a factor of 4. That is, the original material that was compressed to form the Eridanus filaments must have been between 20 and 50 pc in size, depending upon the adopted distances of the filaments. For the age of the filaments to be less than a Myr, as required given the hydrogen recombination time, the original shock must have been travelling at a speed greater than 20 km / s, or greater than 50 km / s if the filaments are farther than 500 pc distant, in order for the shock to have travelled across the entire cloud in less than the recombination time. While this is faster than the current line-of-sight expansion velocity of the superbubble, it is well within the range of plausible expansion speeds for the superbubble. 
	
	Based upon the 25$^\circ$ length of Arc A, the superbubble radius at the location of the Eridanus filaments is between 40 and 110 pc, depending upon the distance of Arc A. If the cloud shocked by the expanding superbubble has been carried along the expanding wall of the superbubble at a speed of the order of 50 km / s, then the initial collision would have had to have occurred roughly 1-2 Myr previously to explain the current radial size of the ring formed by the filaments. This is slightly on the long side if the filaments are to still be recombining from the initial collision. However, this time-scale assumes expansion from the central axis of the superbubble and this time-scale would be reduced if the initial shocked cloud had a significant spatial extent or was offset from the central axis of the superbubble, as hinted at by the hook shape of the filaments.

\section{ORIGIN}
\label{origin}
	
	The origin of the significant column density enhancement in the Eridanus filaments is an open question. The Eridanus filaments may have formed when the superbubble impacted a pre-existing gas cloud and swept up the gas into a dense ring around the outside of the bubble, as suggested above. The original gas cloud that was compressed could have been related to previous star formation events that occurred in one of the older Orion subgroups and the formation mechanism of the Eridanus filaments may bear some similarities to the formation of bipolar rings in planetary nebula and supernova remnants, as such rings are believed to form when a fast outflow impacts a previously ejected shell of material \citep{Soker02}. The apparent thinness of the filaments on the plane of the sky could be partially due to the onset of a thermal instability (e.g., \citealt{Field65}), in which case the filaments would be expected to be slightly cooler than the surrounding wall. 
	
	An alternative possibility is that the breaking out of the ionization front from the superbubble wall at an earlier time may have resulted in a pressure discontinuity that funnelled material into a ring at the height of the filaments, although it is unlikely that such a mechanism could operate on the appropriate time-scale and it is unclear if there was enough material in the top part of the bubble to account for the significant column density of the arcs. Similarly, hydrodynamic instabilities, such as those seen in the simulations of \citet{MacLow89}, can also create structure in a superbubble far from the driving source, but it is not clear how much material can be incorporated in these instabilities. 
	
	The Orion nebula cluster alone has photoevaporated a few times 10$^2$ M$_\odot$ of material over the last megayear \citep{Odell01}. Over the last 10 Myr, the Orion star-forming region should have photoablated between 10$^3$ and a few times 10$^4$ M$_\odot$ of material into the bubble interior and this additional mass injected into the bubble may be the material that has formed the filaments. It is, however, unclear how such photoablated material would become so well focused into filamentary structures.  
	
	The flux of ionizing photons coming from the Orion star-forming region has also likely been quite temporally variable, especially around the occurrences of supernovae. The Eridanus filaments may have formed at a much earlier time and have only recently been ionized by a recent burst of ionizing photons from the Orion star-forming region. Previous supernovae would also have ejected shells of material into the superbubble cavity and the filaments may just be the remnants of previous supernova explosions. More modelling, however, is required for all of the above-suggested possibilities.

\section{CONCLUSIONS}
\label{conclusions}

The Orion star-forming region is the closest high-mass star-forming region currently forming stars and it has blown a large 20$^{\circ}$ by 45$^{\circ}$ superbubble into the ISM. The superbubble contains very prominent filaments on the Eridanus side, referred to as the Eridanus filaments. We find that the Eridanus filaments have gas densities between 1 and 6 cm$^{-3}$ and contain between 300 and 3000 M$_\odot$ of ionized gas. Based upon the widths and H$\alpha$ intensities of these filaments, we find that if these filaments are in ionization equilibrium with the Orion star-forming region, then either the filaments must be edge-on sheets, with depths of a factor of roughly 5 larger than their widths, or that the ionizing luminosity of the Orion star-forming region must be a factor of about 5 larger than previously determined. We suggest, as an alternative, that the Eridanus filaments are non-equilibrium structures that are currently in the process of recombining after being formed from the compression of a pre-existing gas cloud due to the expansion of the superbubble.
    
\section*{ACKNOWLEDGEMENTS}

We would like to thank Dr Basu, Dr Vaidya, and Dr Caselli for useful discussions and comments. AP was partially supported by the Natural Sciences and Engineering Research Council of Canada graduate scholarship programme. DJ acknowledges support from a Natural Sciences and Engineering Research Council (NSERC) Discovery Grant. This research has made use of the Smithsonian Astrophysical Observatory (SAO) / National Aeronautics and Space Administration's (NASA's) Astrophysics Data System (ADS). The WHAM is funded by the National Science Foundation. We would also like to thank our anonymous referee for many useful changes to this paper.

\bibliographystyle{mn2e}
\bibliography{ponbib}{}

\appendix
	
\section{NATURE OF ARC A}
\label{Arc A}

\citet{Reynolds79} were the first to suggest that Arc A might be a part of a large superbubble created by the Orion star-forming region. This interpretation of Arc A being a filament associated with the superbubble, however, has recently come under question, and the precise nature of Arc A is not yet agreed upon (e.g., \citealt{Boumis01}). 

Part of the difficulty in deciphering the nature of Arc A is that there are gas clouds unrelated to the superbubble that also lie along the line of sight towards Arc A. In H {\sc i} and continuum emission, there is significant emission extending from the Galactic plane to a Galactic latitude of approximately -40$^\circ$ along the northern (higher declination) edge of the superbubble (e.g., \citealt{Neugebauer84, Hartmann97, Arnal00, Bajaja05, Kalberla05, MivilleDeschenes05}). In the north-west corner of the loop formed by Arcs A and B, where there is little H$\alpha$ emission, there also lies the end of an 80$^\circ$ long H {\sc i} filament that has a characteristic velocity relative to the local standard of rest (LSR) of -8 km s$^{-1}$, the western half of which is dubbed the Pisces Ridge \citep{Fejes73}. The 100 M$_\odot$, diffuse molecular cloud MBM 18, also known as L1569, is also along the line of sight to the middle of Arc A, near where a linear H$\alpha$ feature, lying along a Galactic longitude of 190$^\circ$, crosses Arc A (e.g., \citealt{Magnani85, Magnani86}). MBM 18 also, unfortunately, has a CO, centroid, LSR velocity between 8 and 10 km s$^{-1}$ \citep{Magnani85, Penprase93,Gir94,Magnani00}, which is similar to the 11 km s$^{-1}$ central velocity of the H$\alpha$ emission coming from Arc A \citep{Reynolds79}.

There are also some weaker CO detections along the northern half of Arc A, at velocities between 7 and 12 km s$^{-1}$, but there are no CO detections along the southern half of Arc A or along Arc B \citep{Magnani00, Aoyama02}. Due to the presence of MBM 18, the Pisces Ridge, and other features present near the northern half of Arc A, we suggest that focusing observational efforts on the southern half of Arc A, where there is little additional CO or 100 micron emission, may provide the best opportunity to constrain the properties of Arc A without foreground or background contamination.

In the following, we review the pertinent arguments for and against the association of Arc A with the Orion--Eridanus superbubble. 

\subsection{Proper motion}
\label{proper motion}

\subsubsection{Against association with the superbubble}
\label{proper motion against}

\citet{Boumis01} derive an upper limit to the proper motion of Arc A of 0.13 arcsec yr$^{-1}$ by tracking the location of a sharp edge of Arc A near the middle of the Arc over a 45 yr time span. If Arc A is 200 pc distant, this proper motion would constrain the tangential motion of the arc to be less than 6 km s$^{-1}$. At distances of 400 and 600 pc, the arc would require tangential velocities less than 11 and 17 km s$^{-1}$, respectively. \citet{Boumis01} assume that the tangential velocity of the bubble is roughly equal to the line-of-sight expansion velocity derived by \citet{Reynolds79}, 15 km s$^{-1}$, and derive a minimum distance to Arc A of 530 pc. Based upon this large distance, they suggest that Arc A might be unassociated with Arc B and the superbubble. 

\subsubsection{For association with the superbubble}
\label{proper motion for}

The conclusions of \citet{Boumis01} are predicated upon the assumption that the radial velocity of Arc A is the same as the velocity of Arc A in the plane of the sky. It is not entirely clear that the tangential velocity of Arc A should be equivalent to the radial expansion velocity, as this assumption essentially prescribes the angle between the space velocity of Arc A and the line of sight. As pointed out in Section \ref{alternative models}, the average expansion velocity of the superbubble must have been much larger than the currently accepted 15 km s$^{-1}$ line-of-sight expansion velocity. Kompaneets model fits to the superbubble \citep{Pon14b} seem to indicate that Arc B may be the better candidate for detecting a proper motion, as the expansion motion of Arc B should be more perpendicular to the plane of the sky than for Arc A. No such proper motion study has yet to be conducted for Arc B.

\subsection{Radial velocities}
\label{velocities}

\subsubsection{Against association}
\label{velocities against}

	 In the \citet{Reynolds79} H$\alpha$ data, two line components are detected with velocities, relative to the LSR, of 3 and -25 km s$^{-1}$. These two components have been interpreted as being the velocities of the near and far side of the superbubble. Similarly, the H {\sc i} lines in the region of the superbubble show complex line structure and are often double peaked. \citet{Menon57} identify one component at a V$_{\mbox{LSR}}$ of 12 km s$^{-1}$ and another at -5 km s$^{-1}$, with these two components again presumably belonging to the two different sides of the superbubble.
	 
	As for the filaments, \citet{Reynolds79} suggest a characteristic velocity of 11 km s$^{-1}$ for Arc A and 3 km s$^{-1}$ for Arc B. In the Leiden/Argentine/Bonn Galactic H {\sc i} Survey data set \citep{Hartmann97, Arnal00, Bajaja05, Kalberla05}, filamentary H {\sc i} emission near Arc A can be seen at velocities from 20 km s$^{-1}$ down to approximately -2 km s$^{-1}$. At velocities between 0 and 10 km s$^{-1}$, the H {\sc i} is slightly offset to the west of Arc A, while between 10 and 20 km s$^{-1}$, the H {\sc i} is essentially coincident with Arc A. This is in contrast to \citet{Reynolds79} who, using the 18 $<$ V$_{\mbox{LSR}} <$ 21 km s$^{-1}$ data of \citet{Heiles76}, place the H {\sc i} filament slightly to the east of Arc A. The long H {\sc i} filament that runs to the west of Arcs B and C has a velocity, relative to the LSR, of 9 km s$^{-1}$ and there also exists a second, fainter H {\sc i} filament further to the west, which is centred closer to 5 km s$^{-1}$ \citep{Verschuur73}. 

	While in both H {\sc i} and H$\alpha$ studies the gas near Arc B has a velocity similar to the larger of the two velocity components detected for the bubble wall, thereby suggesting that Arc B is associated with the far wall of the superbubble, the H {\sc i} and H$\alpha$ emitting gas regions near Arc A have velocities larger than either of the two components detected in H {\sc i} and H$\alpha$ towards the Eridanus side of the bubble. It is unclear what is the cause of this velocity difference. For instance, it may be due to some residual momentum left over from the formation of the arc or to foreground or background material along the line of sight confusing the line centroid from Arc A, but one possible explanation is simply that Arc A is unassociated with the superbubble. 

Taking 20 km s$^{-1}$ as the characteristic velocity for  H {\sc i} gas associated with Arc A, as done by \citet{Welsh05}, implies a distance of 2.2 kpc for the arc, based upon a standard Galactic rotation curve with a circular velocity of 220 km s$^{-1}$. The 10 km s$^{-1}$ velocity associated with the H$\alpha$ emission from Arc A corresponds to a distance of 1.1 kpc. Both of these distances are much larger than the distance to the Orion star-forming region. 

\subsubsection{For association}
\label{velocities for}

\begin{figure}
   \centering
   \includegraphics[height=3in, angle=270]{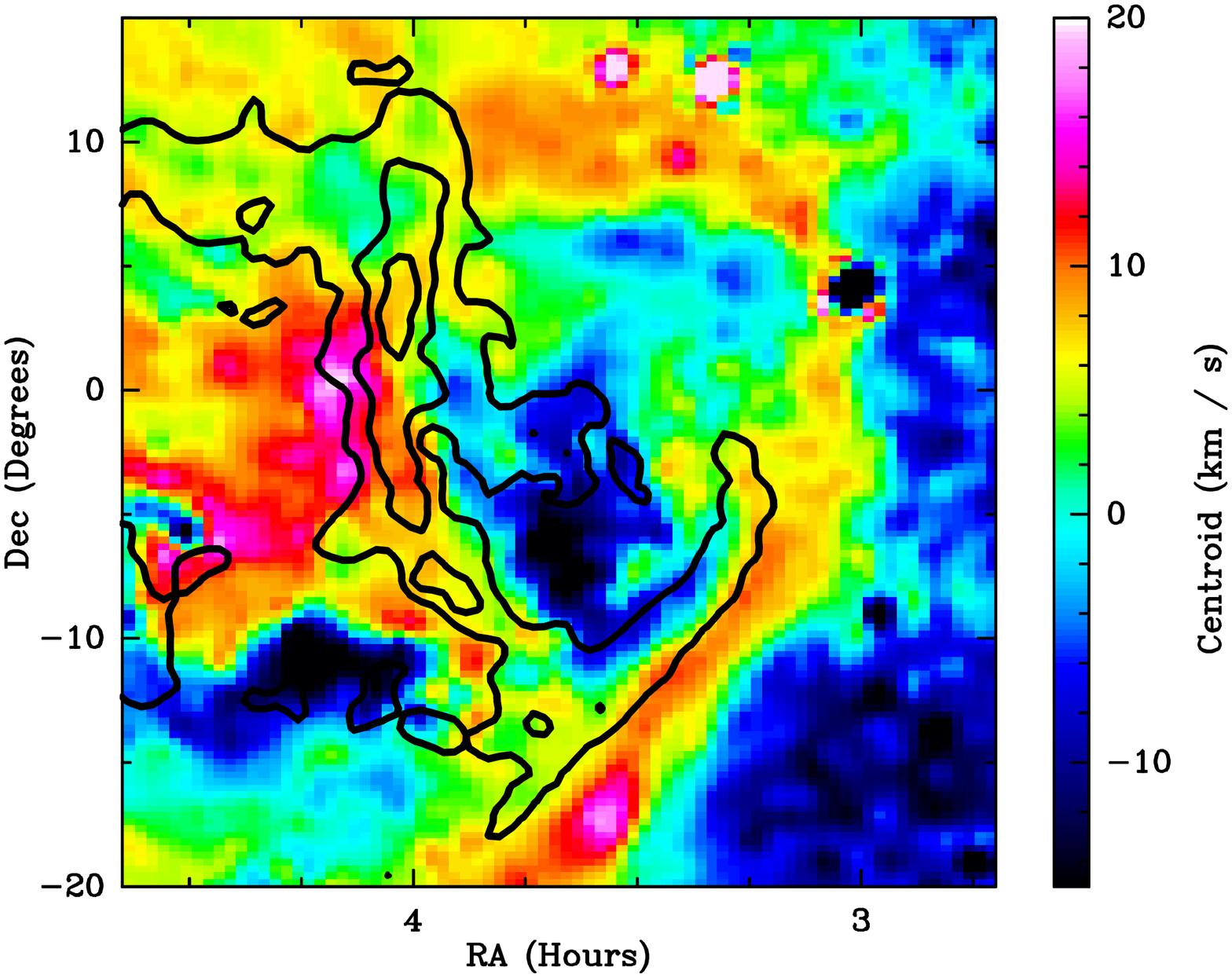}
   \caption{Centroid velocities, with respect to the LSR, of H$\alpha$, from the WHAM survey \citep{Haffner03}, are shown in the colour scale, while the contours show the integrated intensity of the H$\alpha$ line. The contours are logarithmically spaced with each contour representing a factor of 2 increase in integrated intensity. The lowest contour corresponds to an integrated intensity of 10 Rayleighs. Only the region around the Eridanus filaments is shown.}
   \label{fig:whamcent}
\end{figure}

Fig. \ref{fig:whamcent} shows the velocity centroids across the Eridanus filaments, as derived from the WHAM H$\alpha$ data \citep{Haffner03}. The velocity centroids towards Arc A are at slightly more positive velocities than Arc B, although there is a fairly prominent velocity gradient across the two arcs, with the gas along the outside edge of the arcs having centroid velocities more than 10 km s$^{-1}$ larger than the inner edge of the arcs. Arc B is more clearly defined in Fig. \ref{fig:whamcent}, possibly due to the presence of foreground material in the direction of Arc A. The presence of such strong gradients makes it very difficult to accurately assign a particular velocity to the gas in the filaments in order to compare the filament velocities to the bubble wall velocities. 

If the filaments were formed when a pre-existing cloud was shocked and compressed by the expanding superbubble, then the velocity gradients in the filaments could be the kinematic signature left behind by such a collision. It is, however, not clear if such a gradient would be indicative of continuing compression of the filaments or a gradual re-expansion of the filaments. These velocity gradients may also be caused by other physical effects, such as bulk rotation of the filaments.

As noted by \citet{Green93}, because Orion is near $l$ = 180$^\circ$, the motion due to Galactic rotation of any gas cloud towards Orion will be mainly in the plane of the sky and thus, radial velocities do not provide a good measure of distance.  

Arcs A and B appear to join smoothly in velocity space. There is no obvious discontinuity suggesting that the two arcs are located at significantly different distances. 

\subsection{Absorption studies}
\label{absorption}

\subsubsection{Against association}
\label{absorption against}

Towards the Eridanus side of the superbubble, absorption features are seen between -8 and -20 km s$^{-1}$ in stars more distant than 180 pc, which are usually interpreted as being due to the near, approaching side of the superbubble \citep{Guo95, Burrows96, Welsh05}. From colour excesses of stars with known distances, \citet{Lallement14} also find two elongated clouds at a distance of 170 pc from the Sun in the direction of Barnard's Loop.

While the near side of the superbubble is readily detected via absorption features in stellar spectra, there are no unambiguous detections of positive velocity gas with velocities at or above 20 km s$^{-1}$ \citep{Guo95, Burrows96, Welsh05}. Since Arc A is associated with H {\sc i} gas at a velocity of approximately 20 km s$^{-1}$, \citet{Welsh05} argue that Arc A must be more than 500 pc distant. Thus, Arc A would appear to be behind the superbubble. While the back wall of the superbubble has not been conclusively detected in absorption studies \citep{Guo95, Burrows96, Welsh05}, the back side of the bubble has been estimated to be within 540 pc towards ($l$, $b$) = (215$^{\circ}$, -26$^{\circ}$) and within 465 pc towards ($l$, $b$) = (209$^{\circ}$, -37$^{\circ}$) based upon Lyman $\alpha$ to 21-cm ratios \citep{Savage72, Heiles76, Long77} .

\subsubsection{For association}
\label{absorption for}

Towards the superbubble, absorption features are detected around a V$_{\mbox{LSR}}$ of 7 km s$^{-1}$ in the spectra of stars at distances greater than 140 pc and these absorption features are believed to be due to the expanding shell of the local bubble (e.g., \citealt{Frisch90, Burrows96, Lallement03, Welsh05}). This emission is unlikely to be associated with the superbubble, as the gas is moving towards the Orion star-forming region, rather than away from it. Conversely, it is unlikely that the H$\alpha$ emission detected at positive velocities is associated with the local bubble wall, as the local bubble does not contain an obvious ionizing radiation source. 

The near side of the superbubble is detected at velocities between -8 and -20 km s$^{-1}$ in absorption studies (e.g., \citealt{Welsh05}). This velocity range is more positive than the lower of the two H$\alpha$ velocity components, -25 km s$^{-1}$, but more negative than the lower H {\sc i} velocity component, -5 km s$^{-1}$. If the higher of the two detected velocity components, in both H {\sc i} and H$\alpha$, corresponds to the far side of the bubble and if the far bubble wall were to appear in absorption studies at a velocity intermediate to that in H {\sc i} and H$\alpha$ studies, as the near side does, then it would be expected that the far bubble wall would appear at velocities between 3 and 12 km s$^{-1}$. Since the local bubble wall has a velocity of 7 km s$^{-1}$, it is likely that any absorption features due to the far wall of the superbubble would be readily confused with absorption due to the local bubble. Therefore, the lack of a detection of the far bubble wall and Arc A in absorption studies can be simply explained by confusion due to the local bubble wall. Furthermore, the H$\alpha$ centroid of Arc A, 11 km s$^{-1}$, is much lower than the 20 km s$^{-1}$ value adopted by \citet{Welsh05} in their absorption line survey and absorption lines at 11 km s$^{-1}$ could be more readily confused with absorption due to the local bubble or even MBM 18, which has a velocity of 8-10 km s$^{-1}$. \citet{Heiles79} also note that in their catalogue of H {\sc i} shells and supershells, many have only one side detected.

Absorption lines from the back wall of the superbubble would also not be expected to be detected by Na {\sc i} line surveys, such as that by \citet{Welsh05}, if the back wall were fully ionized, because neutral sodium traces gas with temperatures less than 1000K \citep{Lallement03}. 

\subsection{H$\alpha$}
\label{halpha intensity}

\subsubsection{Against association}
\label{halpha intensity against}

There is an abrupt drop in H$\alpha$ intensity to the west of Arc B, along with a corresponding drop in H$_{2}$ emission. \citet{Ryu06} interpret this intensity change as being due to Arc B absorbing most of the UV photons from the Orion star-forming region, although this may also be partly due to a significant decrease in the column density of gas along the line of sight to the west of Arc B. In contrast, the H$\alpha$ and H$_{2}$ intensities immediately to the west of Arc A are not significantly less than the intensities to the east of Arc A. From this, \citet{Ryu06} conclude that Arc A does not lie along the line of sight between the Orion star-forming region and Arc B, as Arc A does not appear to be absorbing all of the UV photons travelling westward from the Orion star-forming region. They note that this is consistent with Arc A being unassociated with the superbubble. 

As suggested above, the H$\alpha$ emission from the Eridanus filaments is not easily compatible with the filaments being in ionization equilibrium with the Orion star-forming region. If the emission is due to the gas recombining, instead of being from reprocessed ionizing radiation, then the filaments do not need to be associated with the Orion superbubble.

\subsubsection{For association}
\label{halpha intensity for}

If Arcs A and B lie along different sides of the bubble, then the requirement that Arcs A and B lie along different sight lines from Orion can also easily be satisfied for models where Arc A is associated with the superbubble.

If the superbubble is split into four quadrants, centred on the Orion star-forming region, then the H$\alpha$ flux is roughly the same in each quadrant, as would be expected if the emission is entirely due to reprocessed ionizing radiation from the Orion star-forming region \citep{Reynolds79}. Not including Arc A's flux would make the Eridanus side of the superbubble underluminous, although flux equality between all of the quadrants would only be expected if the ionizing photons from the Orion star-forming region never break out of the superbubble wall. While the current emission from Arc A may not be fully powered by ionizing photons from Orion, as would be the case if it were out of ionization equilibrium, some source must have initially ionized the gas and it is unclear what this source would be if not the Orion star-forming region.

Arcs A and B appear to join smoothly in the plane of the sky along their southern ends in H$\alpha$ emission. The H {\sc i} filaments tracing these two arcs also appear to connect at their northern ends, although there are numerous potential background and foreground contaminating sources along the north ends of these filaments. If the two arcs are not associated with one another, a more obvious discontinuity between the arcs might be expected. Unlike Arc A, there has been little question in the literature about whether Arc B is associated with the Orion--Eridanus superbubble such that if Arc A is associated with Arc B, it is likely also associated with the superbubble.

\subsection{X-rays}
\label{xrays}

\subsubsection{Against association}
\label{xrays against}

The most prominent difference between the 0.25 and 0.75 keV maps of the Orion--Eridanus superbubble is that the 0.25 keV emission extends south of the intersection of Arcs A and B while the 0.75 keV emission does not \citep{Davidsen72, Williamson74, Naranan76, Long77, Fried80, Nousek82, Singh82, McCammon83, Marshall84, Garmire92, Burrows93, Guo95, Snowden95Burrows, Snowden97}. 

While there seems to be some consensus that the X-ray emission coming from regions with both 0.25 and 0.75 keV emission, dubbed the Eridanus X-ray Enhancement 1 (EXE1) by \citet{Burrows93}, is generated by hot plasma within the bubble, there is some controversy over the nature of the 0.25 keV southern extension, which was named the Eridanus X-ray Enhancement 2 (EXE2) by \citet{Burrows93}. \citet{Brown95} suggest that EXE1 and EXE2 are both related to the superbubble and are not physically separate. The existence of EXE2 provides some evidence that the superbubble might extend south of Arc A, calling into question why an arc associated with the superbubble would stop half way along the bubble. 

While Arc A appears to be associated with a decrease in soft-X-ray emission, the centroid velocities of H$\alpha$ and H {\sc i} emission from Arc A are not compatible with the velocities of the near side of the bubble. Arc A also has a larger characteristic velocity than Arc B, suggesting that Arc A should be farther towards the back side of the superbubble than Arc B, which has velocities consistent with the back side of the bubble. Thus, these velocities seem to suggest that Arc A cannot be on the front side of the superbubble to absorb C-ray's from the interior of the bubble and some other foreground clouds must be causing this absorption. 

\subsubsection{For association}
\label{xrays for}

\citet{Burrows93} consider EXE1 and EXE2 to be physically separate and suggest that EXE2 could be a bubble blown by a single star, although they do not rule out the possibility that EXE2 is a small-scale blowout of the superbubble. \citet{Snowden95Burrows} note that the superbubble is on the edge of a larger scale 0.25 keV enhancement, see also \citet{Snowden95Freyberg}, and they interpret the EXE2 emission to be coming from a diffuse hot halo background. \citet{Heiles99} suggest that the extended 0.25 keV emission is due to hot gas that has leaked out of the superbubble and subsequently cooled slightly. \citet{Heiles99} further point out that no receding portion of the superbubble is detected in either H$\alpha$ or H {\sc i} for locations between Arc A and B and suggest that this is due to there being a hole in the back of the superbubble through which this hot gas can escape. \citet{Burrows93} find that EXE1 is slightly warmer than EXE2, with EXE1 having a temperature of $2.2 \times 10^6$ K and EXE2 having a temperature of $1.5 \times 10^6$ K. If the superbubble wall west of Barnard's Loop is not fully trapping the ionizing radiation from the Orion star-forming region, EXE2 may just be gas heated by radiation that is penetrating the superbubble. All of these options would be consistent with the edge of the superbubble wall running through the intersection point of Arc A and B, as would be expected if Arc A were associated with the superbubble. 

Arc A appears to be strongly associated with a decrease in soft-X-ray emission, even along the southern region of the arc, which would be expected if Arc A were between the Sun and the interior of the superbubble and absorbing the X-rays coming from the hot interior of the bubble (e.g., \citealt{Guo95, Snowden95Burrows, Snowden97}). A model in which Arc A is on the near side of the bubble, at a distance of approximately 180 pc, would be fully consistent with this X-ray absorption. 

\end{document}